\documentclass[aps,prl,preprint,groupedaddress]{revtex4-2}
\usepackage{graphicx}
\usepackage{dcolumn}
\usepackage{bm}
\usepackage{amsmath}
\usepackage{amssymb}

\begin{document}


\title{Ion Heating in Pair-loaded Relativistic Turbulence}

\author{Yusen Wang}
\author{Zhonghai Zhao}
\author{Xinyu Zhu}
\author{Xianxu Jin}
\author{Zhipeng Zhang}
\author{Bin Qiao}

\affiliation{Center for Applied Physics and Technology, HEDPS and State Key Laboratory of Nuclear Physics and Technology, School of Physics, Peking University, Beijing 100871, People’s Republic of China}

\date{\today}

\begin{abstract}
Pair-loaded relativistic turbulence is expected in compact high-energy astrophysical environments and may sustain plasma heating. Using self-consistent three-dimensional radiative particle-in-cell simulations, we identify the mechanism of ion heating mediated by pair-enhanced radiative feedback in relativistic turbulence. We derive the theoretical criterion for the transition to a regime dominated by radiative feedback and show that pair loading promotes the growth and persistence of large, connected optically thick structures. These long-lived structures sustain intense localized ion heating, repeatedly generating prominent hot-ion clumps and substantially enhancing the mean ion temperature. Our results establish a physical link between pair creation and sustained ion heating, providing a theoretical basis for investigating intermittent hot-ion production in black-hole coronae and gamma-ray burst outflows.
\end{abstract}

\maketitle

\newpage

\section{Introduction}
Magnetized turbulence is a potential source of sustained plasma heating in compact high-energy environments, including black-hole coronae and dissipative regions near gamma-ray burst photospheres~\cite{Galeev1979, Zrake2019, Beloborodov2017a, Nattila2022}. 
High-energy observations support models in which continued dissipation supplies energy to compact emitting regions and Compton scattering shapes the high-energy radiation~\cite{Fabian2015, Ricci2018, Ryde2011}. 
X-ray flares and reverberation measurements further reveal the compact and evolving geometry of the emitting coronae~\cite{Wilkins2021, Kara2019}.
Understanding how turbulence sustains ion heating in these environments therefore requires identifying the structures that concentrate energy conversion and the processes that govern their persistence.
Electron–positron pair creation introduces a potentially important feedback by increasing the scattering optical depth and modifying energy exchange between radiation and plasma~\cite{Zrake2019, Fabian2015}.
However, its influence on the spatial localization and persistence of ion heating remains insufficiently understood.

Particle-in-cell (PIC) simulations provide a first-principles kinetic description of how magnetized turbulence heats and accelerates particles~\cite{Comisso2018, Groselj2019, Comisso2024, Mbarek2026}, revealing species-dependent energization~\cite{Zhdankin2019, Comisso2022} and coherent structures inside turbulence~\cite{Ha2025}. Radiative studies of electron–ion turbulence have demonstrated the persistence of hot ions under strong electron cooling~\cite{Zhdankin2021, Nattila2021} and identified shocks and reconnecting current sheets as sites of ion energization~\cite{Groselj2026}. 
Moreover, significant pair loading can also develop in strongly magnetized environments, modifying plasma dynamics and radiative processes~\cite{Lundman2018, Groselj2022}.
Recent simulations of turbulent pair-plasma coronae have established turbulent Comptonization~\cite{Groselj2024} and pair regulation of the plasma state and emitted spectrum~\cite{Nattila2024}.
However, how pair creation reshapes turbulent structures and thereby controls the strength and duration of localized ion heating remains an open question.

In this Letter, we identify the mechanism of ion heating mediated by pair-enhanced radiative feedback in relativistic turbulence using self-consistent three-dimensional radiative PIC simulations. With a theoretical criterion for the transition to a regime dominated by radiative feedback, we show that pair loading promotes the growth and persistence of large, connected optically thick structures, in which efficient and sustained ion heating proceeds through betatron-like energization and substantial conversion of turbulent bulk kinetic energy into internal energy. We find that pair-loaded turbulence thereby repeatedly generates localized hot-ion clumps, acting as a sustained engine of ion heating.

\section{Simulation Methods}
We employ 3D particle-in-Cell (PIC) code WarpX~\cite{WarpX} to self-consistently solve the Vlasov-Maxwell system for a multicomponent plasma. The simulations are performed in a periodic cubic domain of size $R$. The computational grid has spatial resolution $\Delta x=\Delta y=\Delta z=0.5d_{e0}$, where $d_{e0} = \sqrt{m_ec^2/4\pi n_0e^2}$ is the initial electron skin depth. Charged particles inside are assigned an intrinsic escape length $s_{\rm esc}=R/2$. Each particle will be replaced by a new one sampled from a heat reservoir $T_0$ in the local $\mathbf{E}\times\mathbf{B}/B^2$ drifting fluid frame if it diffuses greater than $s_{\rm esc}$ in any direction~\cite{Groselj2024,Gorbunov2025}. A reduced ion-electron mass ratio $\mu=m_i/m_e = 50$ is adopted for computational feasibility. Background photons are continuously injected into the simulation domain following a blackbody spectrum at heat reservoir temperature $T_0$. The photon escape probability is computed according to the local optical depth and photon energy collectively~\cite{Stern1995,Nattila2024}. Prominent quantum electrodynamic (QED) processes are self-consistently incorporated into our PIC simulations using Monte Carlo methods~\cite{Gaudio2020,Groselj2024,Nattila2024}. Particles in each cell are dynamically resampled during the simulation~\cite{Muraviev2021}.

The simulation parameters are specified as follows. The plasma is initially composed of ions and electrons only, with the density $n_0$ chosen to give a global Thomson optical depth 
$\tau_{\rm T}=n_0 \sigma_{\rm T} s_{\rm esc}=1$~\cite{Poutanen2008}. 
The temperature of the heat reservoir is set to be $T_0/m_ec^2=10^{-5}$. The compactness parameter of injected photons is given by $\ell_{\rm ph}\equiv2.7\dot{n}_{\rm ph}R^2T_0\sigma_{\rm T}/m_ec^3=0.01$~\cite{Lightman1987}. 
The strength of the mean magnetic field $\mathbf{B}_0 = B_0 \hat{\mathbf{z}}$ is determined by the ion magnetization parameter $\sigma_i = B_0^2/(4\pi n_0 m_i c^2)$. Turbulence inside the simulation box is driven via "Langevin antenna"~\cite{Tenbarge2014} with strong magnetic fluctuations $\delta B \sim B_0$. The external oscillating current modes are set to be $(k_xR/2\pi, k_yR/2\pi, k_zR/2\pi)=(\pm1,0,\pm1),(0,\pm 1,\pm 1)$, with characteristic frequencies $\omega_0=0.8\times2\pi v_A/R$ and decorrelation rate $\gamma_0=-0.5\omega_0$, where $v_A=c\sqrt{\sigma_i/(\sigma_i+1)}$ is the Alfv\'en speed. Such configurations are identical in all simulation cases to rule out the effect of the magnetic field itself~\cite{Sebastian2025}. The turbulent energy dissipation compactness in the stable state is then~\cite{Stern1995}
\begin{equation}
\ell = \frac{B_0^2 v_A}{8\pi s_0}\frac{R^3\sigma_T}{s_{\rm esc}m_ec^2},
\label{eq1}
\end{equation}
where $s_0$ is the energy-cascade scale. A larger $\ell$ naturally leads to enhanced pair production and we therefore vary the ion magnetization parameter $\sigma_i$ to control the resulting pair loading in our work. By expressing the pair-loading factor $\mathcal{M}_\pm \equiv 2n_+/n_0 $ in terms of $\sigma_i$, we estimate the onset condition of significant pair creation to be $\sigma_{i,\rm th}\simeq 0.62$~\cite{Beloborodov2017, Groselj2026}. Our simulaton system commonly approaches a quasi-steady energy balance after circa $t=3 \sim 5R/v_A$, and energy components then exhibit stable properties with spatio-temporal intermittency shaped by periodic antenna driving, particle escape and QED processes collectively. Additional numerical and theoretical details are included in Supplemental Material.

\section{Results}
\begin{figure}[h]
	\centering
	\includegraphics[width=0.6\columnwidth]{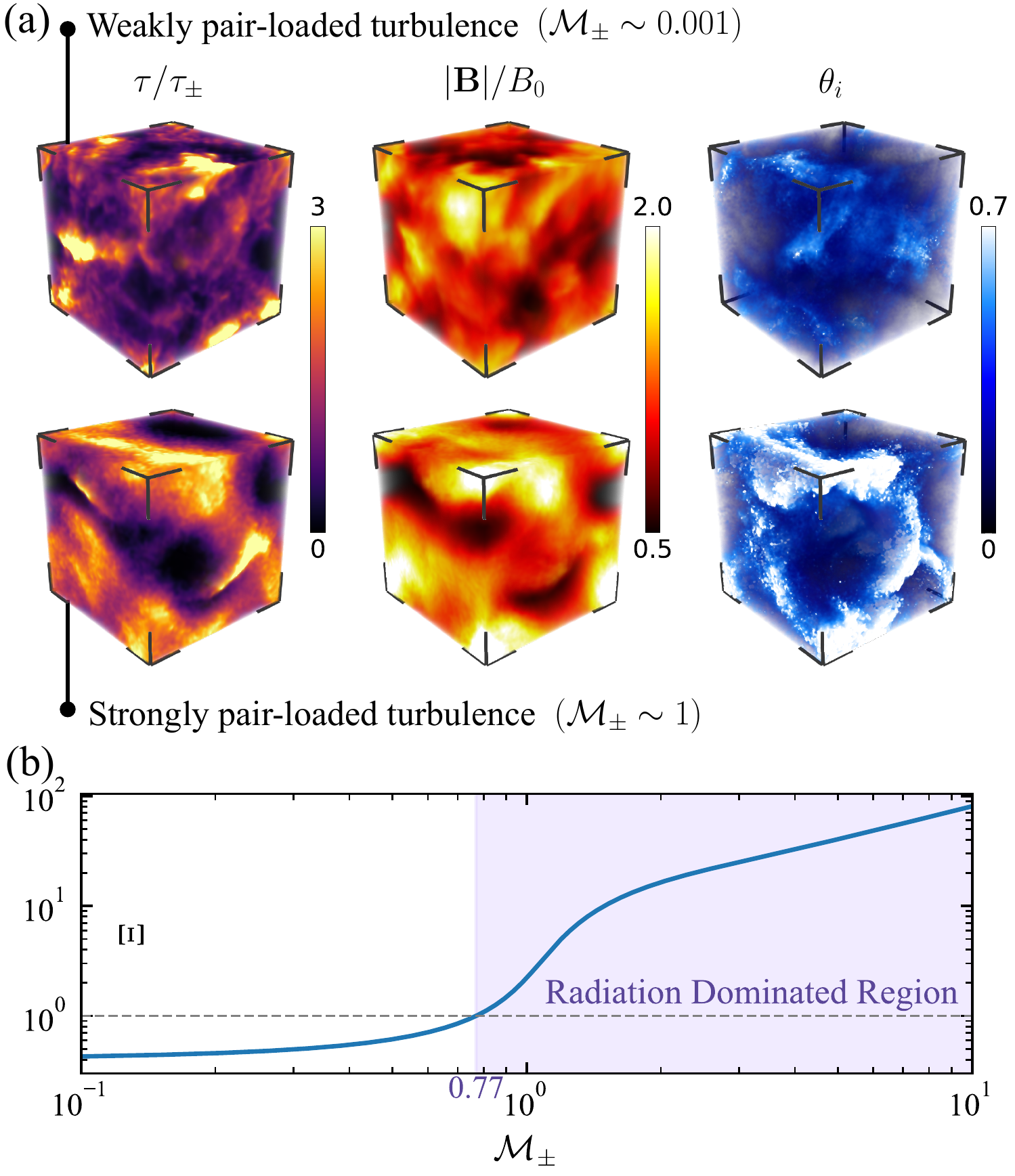}
	\caption{
		(a) Comparison of normalized optical depth $\tau/[\tau_0(1+\mathcal{M}_\pm)] \equiv \tau/\tau_\pm$, magnetic-field strength $|\mathbf{B}|/B_0$, and normalized ion temperature $\theta_i = Ti / m_ic^2$ between weakly (top row) and strongly (bottom row) pair-loaded turbulence.
		(b) Theoretical prediction for $\Xi \equiv \mathcal{R}_\mathrm{rad} / \mathcal{R}_B$ as a function of pair-loading factor $\mathcal{M}_\pm$. Radiative retention becomes significant as increasing higher $\mathcal{M}_\pm$ drives the system into the radiation-dominated regime.
	}
	\label{Fig1}
\end{figure}

In relativistic turbulence, ions serve as the dominant reservoir of turbulent kinetic energy, while leptons remain energetically subdominant as radiative losses rapidly channel their energy into radiation~\cite{Uzdensky2018,Zhdankin2021}. Nevertheless, increased pair loading can substantially modify turbulent dynamics and ion heating through enhanced radiative feedback~\cite{Groselj2022}.
Fig.~\ref{Fig1}a contrasts the distributions of normalized optical depth $\tau/[\tau_0(1+\mathcal{M}_\pm)] \equiv \tau/\tau_\pm$, magnetic-field strength $|\mathbf{B}|/B_0$, and normalized ion temperature $\theta_i = Ti / m_ic^2$ between weakly and strongly pair-loaded turbulence. We show that enhanced pair loading produces extended optically thick structures preferentially associated with enhanced magnetic fields and localized ion heating.

To determine when pair loading modifies the turbulent dynamics, we estimate the threshold at which radiative feedback from pair enrichment overcomes the disruptive effects of magnetic perturbations. 
Intermittent density and magnetic-field structures are continuously distorted by magnetic pressure fluctuations
$\delta P_B \sim f_B \sigma_i n_0 m_i c^2$ in relativistic turbulence~\cite{Vega2022},
where $f_B$ is the effective response factor. The effective enthalpy density for mixed plasma with pair-loading factor $\mathcal{M}_\pm$ gives
$w_\mathrm{eff} \simeq n_0 m_i c^2 [ 1+ (1+\mathcal{M}_\pm)/\mu + 5T_i/2]$. 
We characterize the plasma response to magnetic pressure as
$\mathcal{R}_\mathrm{B} \equiv \delta P_B / w_\mathrm{eff} \sim f_B \sigma_\mathrm{eff}$,
where 
$\sigma_\mathrm{eff}$
is the effective magnetization parameter. The value of $f_B$ can be calculated by relating the turbulence magnetic-pressure response to the dilatational motion over characteristic scale 
$\delta P_B / \ell \sim w_\mathrm{eff} \delta u /c^2 t_{A\ell}$, witch is calculated to be $f_B \simeq 0.13$ in our simulations. 
Enhanced pair-mediated Compton scattering substantially increases the coupling rate between the radiation field and the plasma. The coupling rate is written as 
$\Gamma_c \sim 4 n_0 (1 + \mathcal{M}_\pm)\sigma_{\rm T} c U_\mathrm{ph}/3w_\mathrm{eff}$~\cite{Blumenthal1970,Zrake2019},
and the fraction of radiative coupling achieved within an Alfvénic period is characterized by the saturating form $\eta_c(\mathcal{M}_\pm) = 1 - e^{-\Gamma_c t_{A\ell}}$.
The local photon energy density evolution follows
\begin{equation}
\frac{dU_\mathrm{ph}}{dt} \simeq \frac{U_{B}}{t_{A\ell}} \eta_c(\mathcal{M}_\pm) - \frac{U_\mathrm{ph}}{t_\mathrm{eff}},
\label{eq2}
\end{equation}
where $t_\mathrm{eff} \sim (1 + \tau_\pm) \ell / c$ 
is the effective photon escape time. In the quasi-steady state we yield
$U_\mathrm{ph}/U_B \simeq t_\mathrm{eff} \eta_c(\mathcal{M}_\pm) / t_{A\ell}$.
For a locally quasi-isotropic photon field, the radiation stress effectively coupled to the plasma can be approximated as
$\delta P_\mathrm{rad} \sim g(\tau_\pm) U_\mathrm{ph} / 3$,
where $g(\tau_\pm) = \tau_\pm / (1 + \tau_\pm)$ 
smoothly interpolates between the optically thin, weak-coupling regime and the saturated-coupling limit. We hereby define
$\Xi \equiv \mathcal{R}_\mathrm{rad} / \mathcal{R}_B$ 
to quantify the relative importance of radiative retention and magnetic-pressure-driven turbulent perturbations. Combining aforementioned relations yields an implicit equation for $\Xi$ (see Supplemental Material for derivation)
\begin{equation}
		\Xi ={}
		\Xi_0 + \frac{\tau_\pm}{3 f_B}
		\sqrt{\frac{\sigma_\mathrm{eff}}{1 + \sigma_\mathrm{eff}}}
		\times
		\left[
		1-\exp\left(
		- 8 f_B [1 + \tau_\pm] \sqrt{(1 + \sigma_\mathrm{eff}) \sigma_\mathrm{eff}}
		\frac{\ell}{R} \Xi
		\right)
		\right],
	\label{eq3}
\end{equation}
which is theoretically determined by pair-loading factor $\mathcal{M}_\pm$ shown in Fig.~\ref{Fig1}b. As $\mathcal{M}_\pm \sim 0.77$, $\Xi$ approaches and exceeds unity, indicating that radiative feedback becomes sufficiently strong to preserve the optical-depth enhancement of compressed structures against turbulent decorrelation. We therefore identify $\Xi \gtrsim 1$ as the radiation-dominated regime, in which the turbulent dynamics are qualitatively altered.

\begin{figure}[h]
	\centering
	\includegraphics[width=\textwidth]{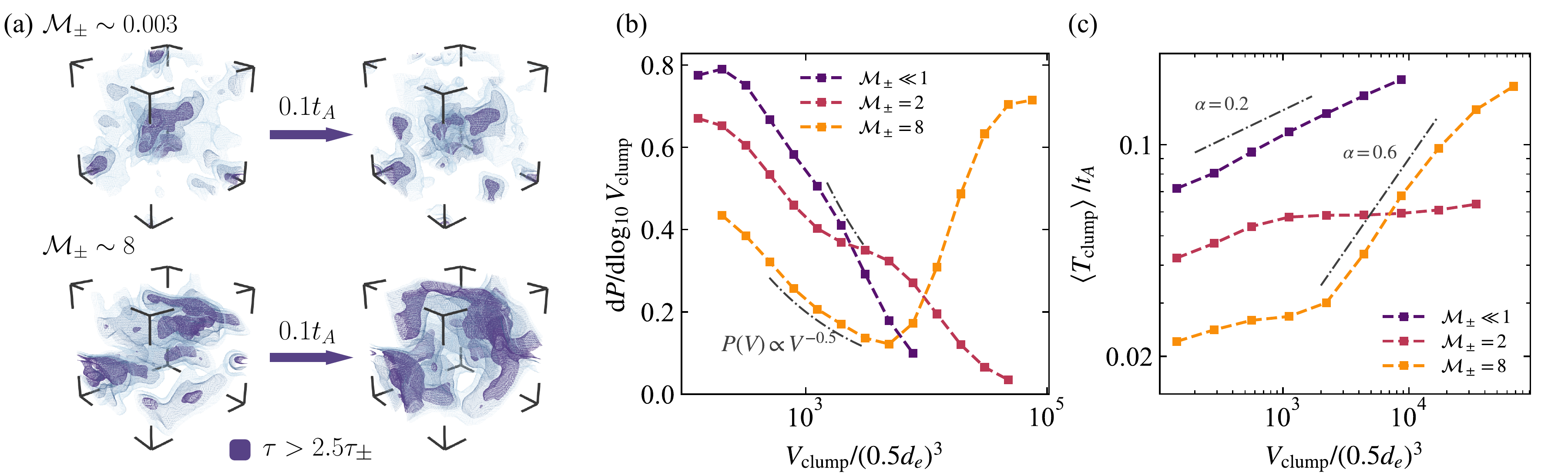}
	\caption{
		(a) Evolution of optically thick regions with $\mathcal{M}_\pm= 0.003$ and $\mathcal{M}_\pm = 8$ within 0.1$t_A$. Clump cores are defined by $\tau > 2.5\tau_\pm$ in each case.
		(b) Clump-volume probability density $\mathrm{d}P/\mathrm{d}\log_{10}V_{\rm clump}$. Changes in the distributions reveal a transition from isolated clumps produced by local compression to large connected structures sustained by radiative retention as pair loading increases.
		(c) Characteristic clump lifetime as a function of volume. Scattering-driven positive feedback promotes the incorporation of small clumps into larger connected structures, steepening the lifetime–volume scaling from $T\propto V^{0.2}$ to $T\propto V^{0.6}$.
	}
	\label{Fig2}
\end{figure}

Based on theoretical analysis, we select representative cases with $\mathcal{M}_\pm= \{0.003, 8\}$ and contrasts the evolution of high optical depth regions in Fig.~\ref{Fig2}a. Under low pair loading condition, optically thick regions arise from local compression driven by motions with sound speed $c_s \sim v_A$, remaining spatially isolated, short-lived due to magnetic perturbations. In contrast, as increasing magnetization raises the pair-loading factor to radiation dominated region, radiative feedback enables optically thick clumps to resist perturbations and persist locally as our theory predicts. Long-lived clumps subsequently trigger positive feedback loop: low-energy photons are prominently upscattered above $m_e c^2$, and then annihilate to produce electron–positron pairs, further increase the scattering opacity. Such feedback then prompt initially isolated regions to merge into a connected structure approaching the box scale, until the balance is disrupted.

To quantify the temporal evolution of high-optical-depth clumps sustained by radiative feedback, we track their formation, disappearance, mergers, and splits~\cite{Zhdankin_2015a, Zhdankin_2015b}, then measure their characteristic sizes and lifetimes across different pair-loading conditions. The tracking criteria are detailed in the Supplemental Material. 
We analyze three representative cases over extended periods and identify optically thick clumps using the identical threshold $\tau>2.5\tau_\pm$. The resulting clump-volume distributions
$\mathrm{d}P/\mathrm{d}\log_{10}V_{\rm clump}$,
directly reflect the coalescence driven by the positive feedback described above [Fig.~\ref{Fig2}b]. In general, clumps produced solely by local compression approximately follow $P(V)\propto V^{-0.5}$. As $\mathcal{M}_\pm$ increases, the distribution extends to larger volumes, with sustained coalescence producing multiple connected structures with $V>10^{3}d_e^{3}$. By $\mathcal{M}_\pm = 8$, clumps predominantly exist as large-scale type rather than intermittent fragments. Considering the corresponding clumps' lifetime, larger structures are commonly more resistant to disruption, with scaling approximately as $T(V) \propto V^{0.2}$. Higher $\mathcal{M}_\pm$ forces small structures to incorporate into larger clumps while retaining the same lifetime–volume scaling. The resulting large connected structures exhibit enhanced persistence, with the scaling steepening to $T(V) \propto V^{0.6}$, and providing a stable environment for sustained scattering-driven positive feedback and other energy-exchange processes.

\begin{figure}[h]
	\centering
	\includegraphics[width=0.6\columnwidth]{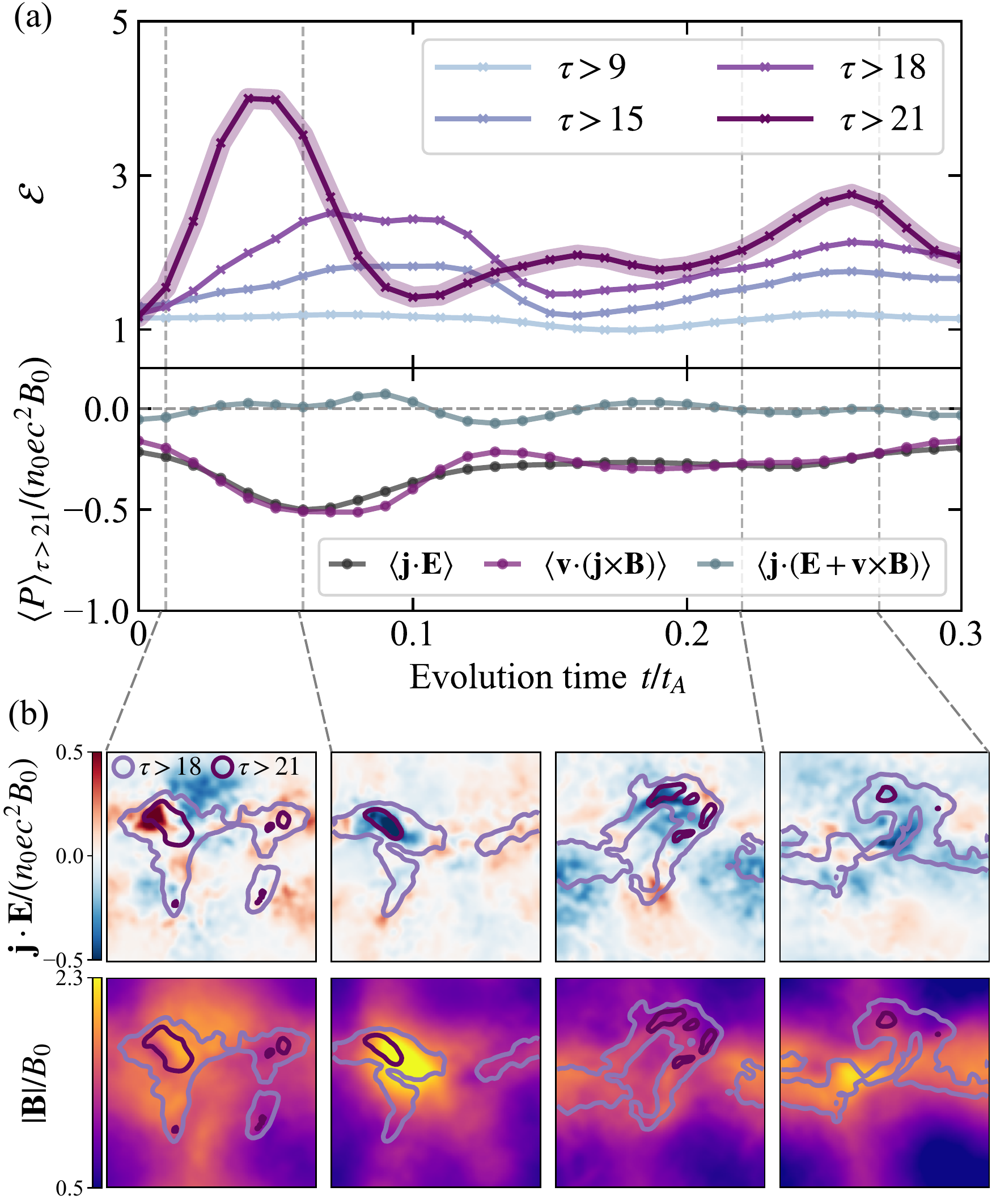}
	\caption{
		(a) Evolution of the overlap enhancement at different optical depths $\tau$, together with the power densities associated with work against the Lorentz force $-\mathbf{v}\cdot(\mathbf{j}\times\mathbf{B})$ and nonideal energy conversion $\mathbf{j}\cdot(\mathbf{E}+\mathbf{v}\times\mathbf{B})$, within regions satisfying $\tau>21$ over an interval of $0.3t_A$. Energy transfer to the magnetic field occurs preferentially in optically thick clumps and is dominated by work against the Lorentz force.
		(b) Cross-sectional snapshots of coherent structures in $\mathbf{j}\cdot\mathbf{E}$ and the normalized magnetic-field strength during two magnetic energization episodes. Intense energy exchange occurs in regions of high optical depth.
	}
	\label{Fig3}
\end{figure}

To examine the physical processes underlying clump evolution, we analyze a \(0.3t_A\) interval encompassing the snapshot in Fig.~\ref{Fig2}a for the $\mathcal{M}_\pm=8$ run.
Optically thick, radiation-loaded clumps serve as sites of sustained plasma-to-field energy conversion. We quantify the spatial localization of strong negative electromagnetic work through the overlap enhancement
\begin{equation}
	\mathcal{E} = \frac{P(\mathbf{j} \cdot \mathbf{E} < -(jE)_\mathrm{rms}, \tau > \tau_\mathrm{th})}{P(\mathbf{j} \cdot \mathbf{E} < -(jE)_\mathrm{rms}) P(\tau > \tau_\mathrm{th})}.
\end{equation}
Here, $\mathcal{E}=1$ denotes statistical independence, whereas $\varepsilon>1$ indicates preferential plasma-to-field energy transfer within  $\tau_\mathrm{th}$. 
The enhancement becomes more pronounced at higher optical-depth thresholds, reaching \(\mathcal{E}\simeq4\) for \(\tau>21\) [Fig.~\ref{Fig3}a], indicating the preferential occurrence of coherent regions within optically thick clumps. Within the sampled interval, two episodes of spatially coherent plasma-to-field energy transfer develop, accompanied by local magnetic-field amplification depicted in Fig.~\ref{Fig3}b. To identify the dominant energy-transfer channel, we decompose the electromagnetic work by
$\mathbf{j} \cdot \mathbf{E}= \mathbf{v}\cdot (\mathbf{j} \times \mathbf{B}) + \mathbf{j} \cdot (\mathbf{E} + \mathbf{v} \times \mathbf{B})$, and evaluate both contributions within $\tau>21$. As shown in [Fig. 3(b)], the relativistic flow transfers bulk kinetic energy back into the electromagnetic field primarily by doing work against the Lorentz force, while nonideal contribution remains secondary. 
An individual episode persists for nearly \(0.2t_A\), substantially longer than instantaneous energy dissipation. The increased effective inertia and strong radiative coupling $\Gamma_c t_A\gg1$ associated with pair loading prolong localized magnetic energization and maintain strong-field regions in which substantial ion heating can therefore emerge.

Localized hot-ion clumps develop preferentially within the strongly magnetized regions delineated by the $|\mathbf{B}|/B_0 > 1.5$ contours in Fig.~\ref{Fig4}a. The detailed spatiotemporal evolution is presented in the Supplemental Material. To separate ion thermal motion from the relativistic bulk flow, we construct the ion pressure tensor in the local Eckart frame~\cite{Eckart1940}. Lorentz transforming the ion energy-momentum tensor gives
$P_{i,E}^{jk} = \Lambda^j_{\mu} \Lambda^k_{\nu} T_{i}^{\mu \nu}$.
Its projections along and across the comoving magnetic field 
$\mathbf{B_E}$
yield 
$T_{i\parallel} = \mathbf{b} \cdot \mathbf{P}_{\rm i,E} \cdot \mathbf{b} / n_i$ and
$T_{i\perp} = (\mathrm{Tr}\mathbf{P}_{\rm i,E} - n_iT_{i\parallel}) / 2n_i$,
where $n_i$ is the proper ion number density and 
$\mathbf{b} = \mathbf{B_E} / |\mathbf{B_E}|$.
We then normalized them by $\theta_{i\parallel}=T_{i\parallel}/m_ic^2$ and $\theta_{i\perp}=T_{i\perp}/m_ic^2$.
Pronounced temperature anisotropy develops in strongly magnetized regions with $\theta_{i\perp} / \theta_{i\parallel} > 5$ inside hot-ion clumps. Such directional preference is consistent with betatron-like energization as the field strengthens. Moreover, the hot regions simultaneously exhibit enhanced $\theta_{i\perp}/|\mathbf{B}_E|$ by several times. This suggests an additional nonadiabatic heating contribution other than magnetic amplification~\cite{Bacchini2022}. Below, we quantify this contribution through the work of nongyrotropic ion stresses against turbulent velocity gradients.

\begin{figure}[t]
	\centering
	\includegraphics[width=0.6\columnwidth]{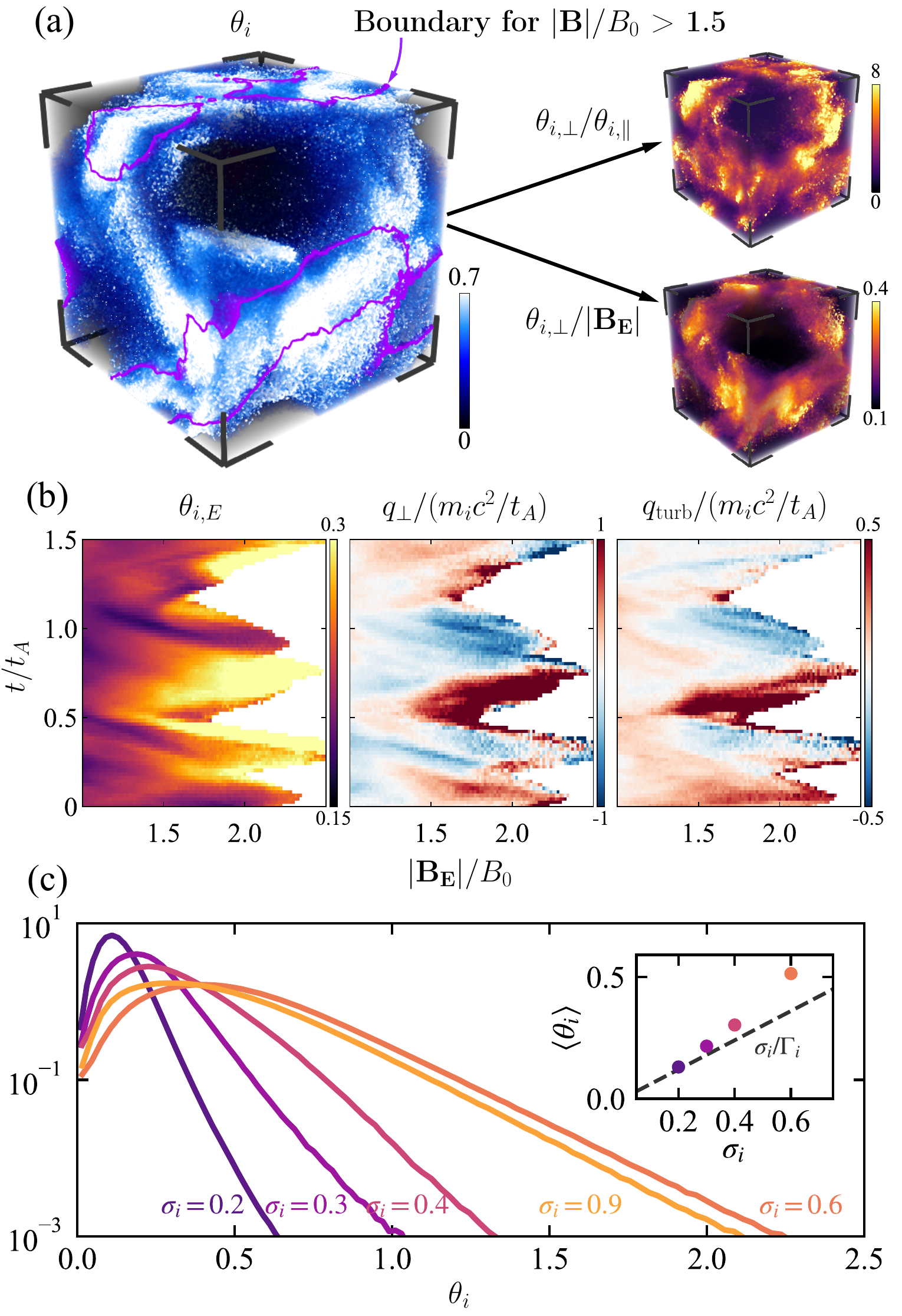}
	\caption{
		(a) Snapshot of the three-dimensional ion-temperature distribution $\theta_i$ for $\mathcal M_\pm =8$. strongly magnetized regions are delineated by the $|\mathbf{B_E}|/B_0 > 1.5$ contours. Pronounced anisotropy and nonadiabatic heating are also observed.
		(b)  Long-term evolution of the ion temperature in the Eckart frame $\theta_{i,E}$, and the perpendicular gyrotropic and nongyrotropic work rates per ion, binned by $|\mathbf{B}|/B_0$. Recurrent magnetic amplification and conversion of bulk kinetic energy into thermal energy jointly drive ion heating.
		(c) Ion-temperature distributions for different $\sigma_i$. The insets show deviations from the theoretical prediction caused by heating in pair-loaded turbulence.
	} 
	\label{Fig4}
\end{figure}

We examine two contributions to ion energization using a three-dimensional pressure–strain diagnostic. Decomposing the pressure tensor as
$
\mathbf P_{i,E}
=P_\perp\mathbf I+(P_\parallel-P_\perp)\mathbf{bb}
+\boldsymbol{\Pi},
$
we evaluate the perpendicular gyrotropic and nongyrotropic work rates per ion
\begin{equation}
q_\perp
=-\frac{P_\perp}{n_i}
\left(\nabla\cdot\mathbf v_i-\mathbf{bb}:\nabla\mathbf v_i\right),
\:\:q_{\rm turb}
=-\frac{\boldsymbol{\Pi}:\mathbf S_i}{n_i},
\end{equation}
where $S_{ij}=(\partial_i v_j + \partial_j v_i)/2-\delta_{ij}\nabla\cdot\mathbf{v}/3$
is the traceless rate-of-strain tensor.
The perpendicular contribution measures pressure work associated with deformation across the local magnetic field and captures betatron-like energization in the magnetized, adiabatic limit. The nongyrotropic contribution measures the coupling of ion stresses to turbulent shear and deformation, with positive values corresponding to ion energization.

Ion energization remains recurrent and intermittent throughout the $1.5t_A$ interval, with elevated ion temperatures in Eckart frame $\theta_{i,E}$ repeatedly developing in the amplified-field regions depicted in Fig.~\ref{Fig4}b. Within the magnetic structures, perpendicular pressure work reaches
$q_\perp=O(m_ic^2/t_A)$, supporting rapid ion energization over the structures’ lifetime. Meanwhile, nongyrotropic pressure work is strongly enhanced in the same high-field range, with $q_{\rm turb}$ approaching \(q_\perp\) in magnitude. This substantial contribution couples turbulent bulk kinetic energy to ion internal energy and augments ion heating to higher level. The persistence of coherent structures allows both channels to operate progressively and support hot-ion clumps formation inside relativistic turbulence. Energization within each clump is nevertheless finite: heated ions leave the domain through the prescribed escape process, while the evolving balance between magnetic and plasma stresses allows the strong-field structures to relax. The resulting recurrent process of structure formation, energization and relaxation constantly produces hot-ion clumps and supplies hot ions to the escaping population.

Fig.~\ref{Fig4}c shows the ion-temperature distributions for different $\sigma_i$. Increasing $\sigma_i$ substantially broadens the distributions and extends their high-temperature tails, which mostly embedded in the hot-ion clumps described above.  Stronger pair loading raises the mean ion temperature $\langle\theta_i\rangle$ substantially above theoretical prediction in the absence of pair loading $\Gamma_i\sigma_i$(Fig.~\ref{Fig4}c inset)~\cite{Groselj2026}, enhancing the overall ion heating efficiency. Notably, although $\mathcal M_\pm$ reaches 8 at $\sigma_i=0.9$, escape model causes the large-scale hot-ion clumps to disperse before they are fully heated, limiting the temperature reached in this run. Larger-domain simulations are therefore needed to explore the upper limit of ion heating in strongly pair-loaded plasmas in the future.

\section{Conclusions}
In summary, our self-consistent three-dimensional radiative particle-in-cell simulations reveal how pair creation reshapes relativistic turbulence and enables recurrent, localized ion heating. 
Our theoretical analysis determines the critical pair-loading factor $\mathcal{M}_\pm$ at which turbulence enters a regime dominated by radiative feedback for the parameters considered. 
In this regime, sustained positive feedback between Compton scattering and pair creation promotes the merger of optically thick clumps into large, persistent structures. Plasma flows inside continually work against the Lorentz force, efficiently transferring kinetic energy into the electromagnetic field and sustaining localized magnetic amplification. 
Substantial ion heating results from the combined action of betatron-like energization and conversion of turbulent bulk kinetic energy into ion internal energy, with comparable contributions diagnosed by $q_\perp$ and $q_{\rm turb}$. The recurrent formation and heating of hot-ion clumps raise the mean ion temperature above the expectation without pair loading, making pair-loaded turbulence a sustained source of hot-ion clumps potentially available for outward transport.

Our results connect pair-mediated radiative feedback to the spatial organization of ion heating. The mechanism provides a potential route to intermittent hot-ion production in compact, magnetized radiation sources containing baryons, including black-hole coronae and dissipative regions near or below the photospheres of gamma-ray burst jets. Its relevance to these environments can be assessed within our theoretical framework through the criterion $\Xi\gtrsim1$.
Larger-domain simulations with more realistic escape treatments are needed in the future to determine the attainable ion temperatures, the subsequent transport and survival of hot-ion clumps.

\bibliographystyle{apsrev4-2}
\bibliography{citations}


\clearpage

\setcounter{section}{0}
\setcounter{equation}{0}
\setcounter{figure}{0}
\setcounter{table}{0}

\renewcommand{\thesection}{S\arabic{section}}
\renewcommand{\theequation}{S\arabic{equation}}
\renewcommand{\thefigure}{S\arabic{figure}}
\renewcommand{\thetable}{S\arabic{table}}

\section*{Supplemental Material}

\section{Quasi-Steady Evolution and Threshold for Significant Pair Creation }
\begin{figure}[h]
	\centering
	\includegraphics[width=\columnwidth]{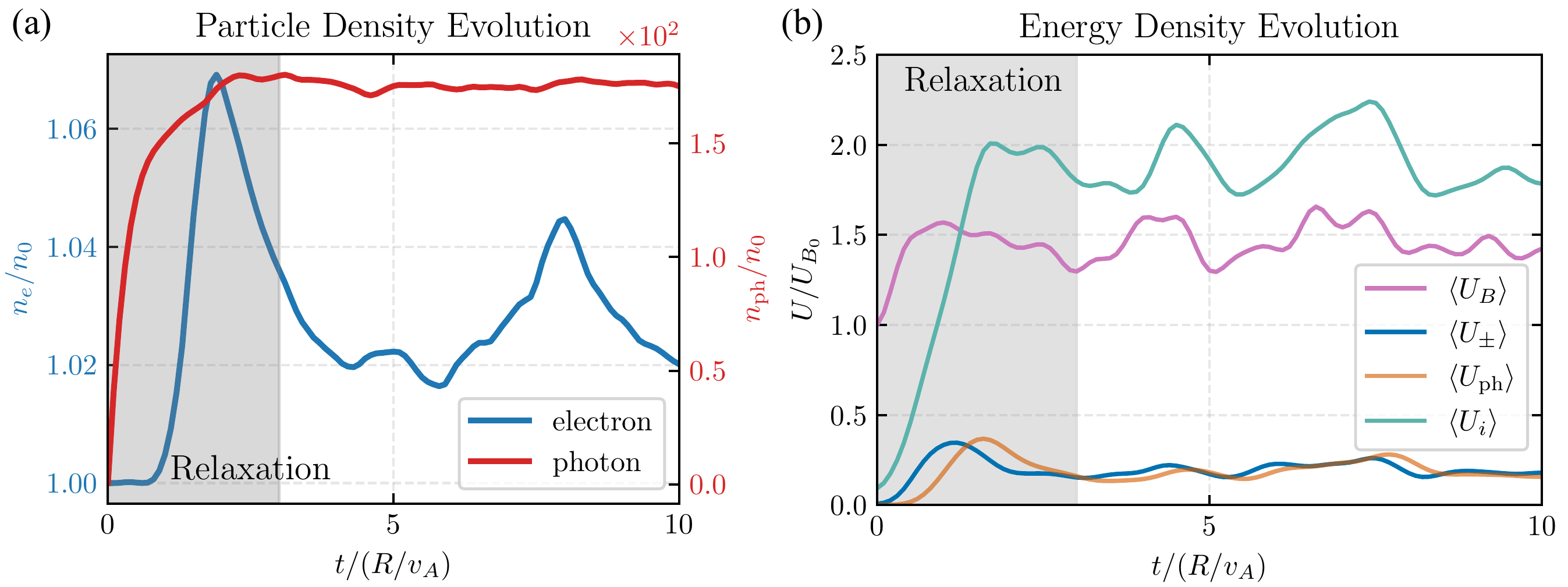}
	\caption{
		(a) Evolution of normalized electron and photon number density. (b) Evolution of normalized energy densities of pairs ($U_\pm$), ions ($U_i$), photons ($U_{\rm ph}$) and magnetic fields ($U_B$).
	}
	\label{FigS1}
\end{figure}

Following an initial relaxation phase about $t\simeq (3$–$5)R/v_A$, the system approaches a statistically quasi-steady state characterized by nearly stationary mean energy densities and total particle number with persistent spatiotemporal intermittency. Fig~\ref{FigS1} shows the evolution of normalized electron and photon number density, together with the normalized energy densities of pairs ($U_\pm$), ions ($U_i$), photons ($U_{\rm ph}$) and magnetic fields ($U_B$) in our fiducial simulation. At the condition of relatively low ion magnetization $\sigma_i=0.2$, pair creation remains subdominant, with a mean $\mathcal{M}_\pm\simeq 0.003$. The subsequent statistical stationarity provides a well-defined regime for analyzing the temporal evolution of dissipative structures in relativistic turbulence.

We estimate the critical ion magnetization required for substantial pair loading following Beloborodov~\cite{Beloborodov2017} and Gro\v{s}elj et al.~\cite{Groselj2026}. Assuming steady-state radiative energy conservation, we obtain
$(\bar{E}_{\rm esc}-\bar{E}_0)\dot{n}_{\rm ph}=(1-p_i)B_0^2v_A/8\pi s_0$,
where $p_i$ denotes the fraction of dissipated energy allocated to ion heating and particle escape, while $\bar{E}_{\rm esc}$ and $\bar{E}_0$ are the mean energies of escaping and background photons, respectively. Meanwhile, balancing pair creation and annihilation yields
$0.1 n_H^2 \simeq 3n_+(n_0+n_+)/8$,
where $n_H=f_Hn_{\rm ph}$ and $f_H$ denotes the fraction of photons with $h\nu>m_ec^2$. The condition for significant pair loading $\mathcal{M}\pm\gtrsim 1$ corresponds approximately to $f_Hn{\rm ph}/n_0\gtrsim\sqrt{3}$. Combining these relations gives the threshold ion magnetization,
\begin{equation}
	\sigma_i \geq \left[ \sqrt{3} \left(\frac{\dot{n}_{\rm ph}}{n_{\rm ph}}\right) \frac{R(\bar{E}_{\rm esc}-\bar{E}_0)}{m_ic^3(1-p_i)}\cdot \frac{1}{f_h} \right]^{\frac{2}{3}}.
\end{equation}
For our fiducial simulation, we measure $\bar{E}_{\rm esc}/\bar{E}0\simeq312$, $p_i\simeq0.555$, and $f_H\simeq1.66\times10^{-3}$, yielding a critical magnetization of $\sigma_{i,\rm th}\simeq0.62$ for substantial pair creation.

\section{Tracking Criteria for optically thick clumps during evolution}

We identify and track optically thick clumps in each case to characterize their sizes and lifetimes. To ensure statistical robustness, in practical, we track the system evolution over a period longer than $10t_A$ and successive snapshots are separated by $8 \Delta t_{\rm CFL}$. At each snapshot, we select cells satisfying 
$\tau(\mathbf{x},t) / \tau_0 (1 + \mathcal{M}_\pm) > 1.5$. Selected cells are grouped into connected structures using a three-dimensional 26-neighbor criterion~\cite{Zhdankin_2015a, Zhdankin_2015b}. Periodic boundaries are included when determining connectivity, and structures containing fewer than 200 diagnostic-grid cells are discarded. Two structures in consecutive snapshots are linked if at least one cell in the earlier structure overlaps with, or lies within the 26-neighbor stencil of a cell in the later structure. These links define the predecessors and successors of each structure and allow for both mergers and splits. Structure identities are assigned according to the following rules.

\begin{figure}[h]
	\centering
	\includegraphics[width=\textwidth]{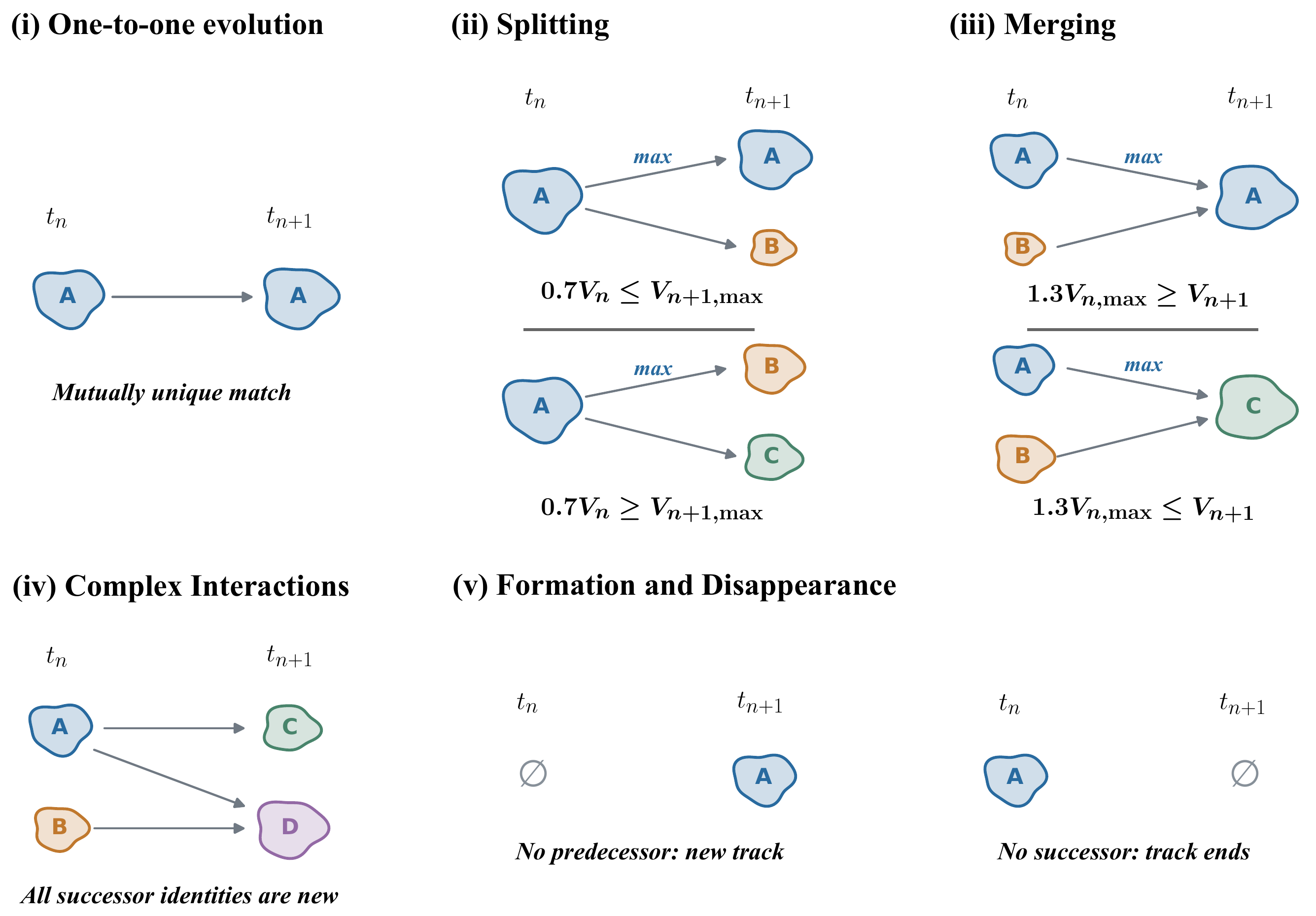}
	\caption{Illustration of tracking criteria}
	\label{fig:S3}
\end{figure}

(i) \textbf{One-to-one evolution}. If a structure has exactly one successor and that successor has exactly one predecessor, the successor inherits the original identity.

(ii) \textbf{Splitting}. A pure split occurs when a structure has multiple successors, each of which has only that structure as its predecessor. The largest successor retains the original identity if its volume satisfies $V_{\mathrm{child,max}}\geq0.7V_{\mathrm{parent}}$; all other successors receive new identities. If this condition is not met, the original track terminates and all successors initiate new tracks.

(iii) \textbf{Merging}. A pure merger occurs when multiple structures share a single successor and each predecessor has only that successor. The merged structure may inherit a predecessor's identity if $V_{\mathrm{new}}\leq1.3V_{\mathrm{parent}}$. If several predecessors satisfy this condition, the largest is selected. The tracks of all other predecessors terminate. If none satisfies the condition, the merged structure initiates a new track.

(iv) \textbf{Complex interactions}. When splitting and merging are intertwined, no identity is inherited across the interaction: the participating predecessor tracks terminate, and the successor structures initiate new tracks.

(v) \textbf{Formation and disappearance}. A structure without a predecessor initiates a new track, whereas a structure without a successor terminates its track.

For each track, we define the lifetime as
$T_{\mathrm{clump}}=t_{\mathrm{last}}-t_{\mathrm{first}}$
where $t_{\mathrm{first}}$ and $t_{\mathrm{last}}$ are its first and last detected times. Its characteristic volume is the mean over the uniformly spaced snapshots along the track,
$\overline{V}=\frac{1}{N_{\mathrm{snap}}}\sum_{k=1}^{N_{\mathrm{snap}}}V_k,$
where $V_k$ is the number of constituent diagnostic-grid cells. We exclude tracks already present at the first snapshot or still present at the last snapshot, since their full lifetimes are not observed. Tracks detected in only one snapshot are also excluded because their lifetimes are unresolved at the adopted cadence. The resulting statistics describe the persistence of connected high-optical-depth structures under these tracking criteria, rather than the residence time of individual plasma elements.

\section{Spatiotemporal formation of hot ion clumps}

\begin{figure}[h]
	\centering
	\includegraphics[width=\textwidth]{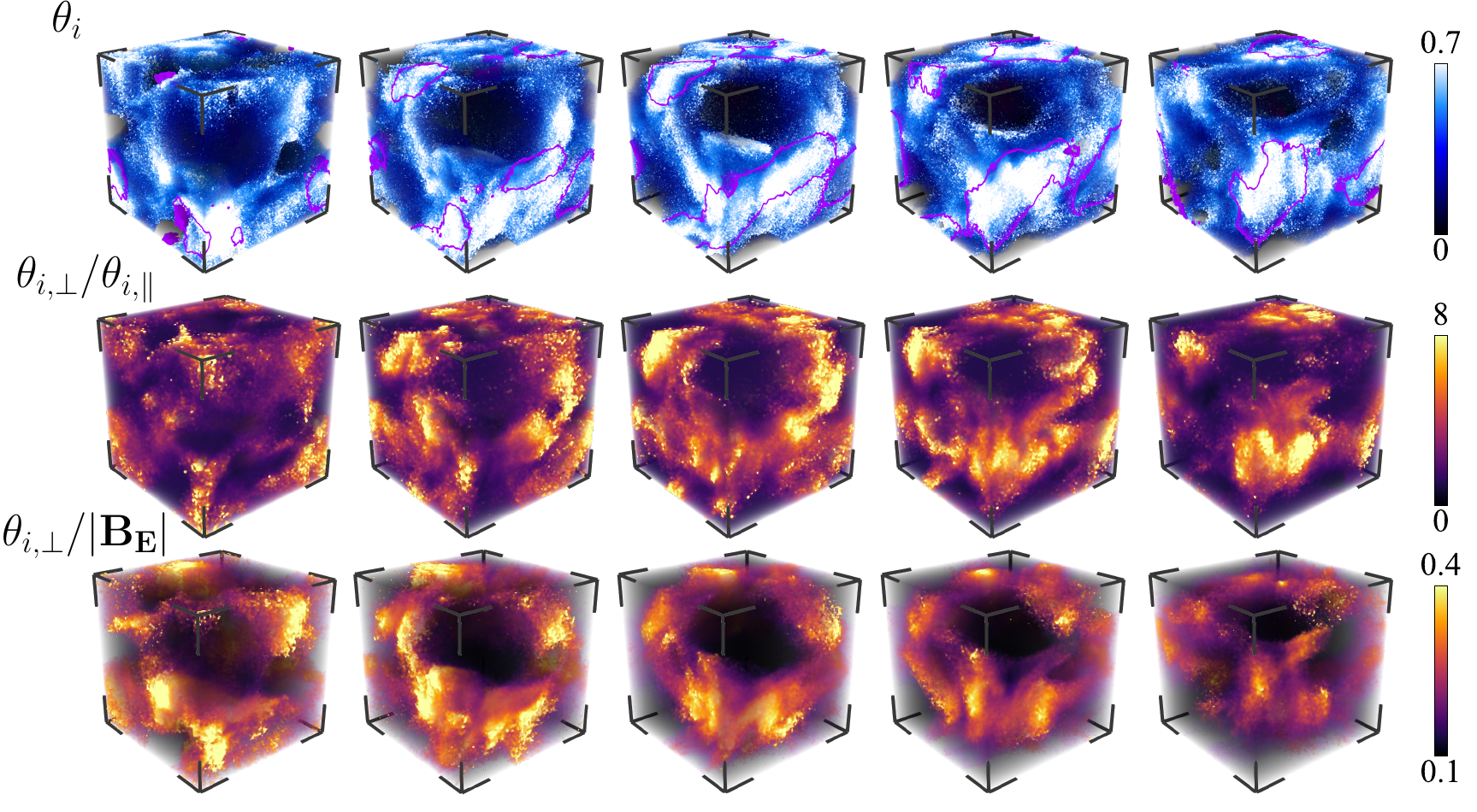}
	\caption{Detailed spatiotemporal formation of hot ion clumps}
	\label{FigS4}
\end{figure}

Fig~\ref{FigS4} shows the detailed spatiotemporal formation of hot ion clumps within growing regions of strong magnetic field, followed by their gradual escape from the simulation domain and dispersal, accompanied by pronounced anisotropy.

\begin{figure}[h]
	\centering
	\includegraphics[width=\textwidth]{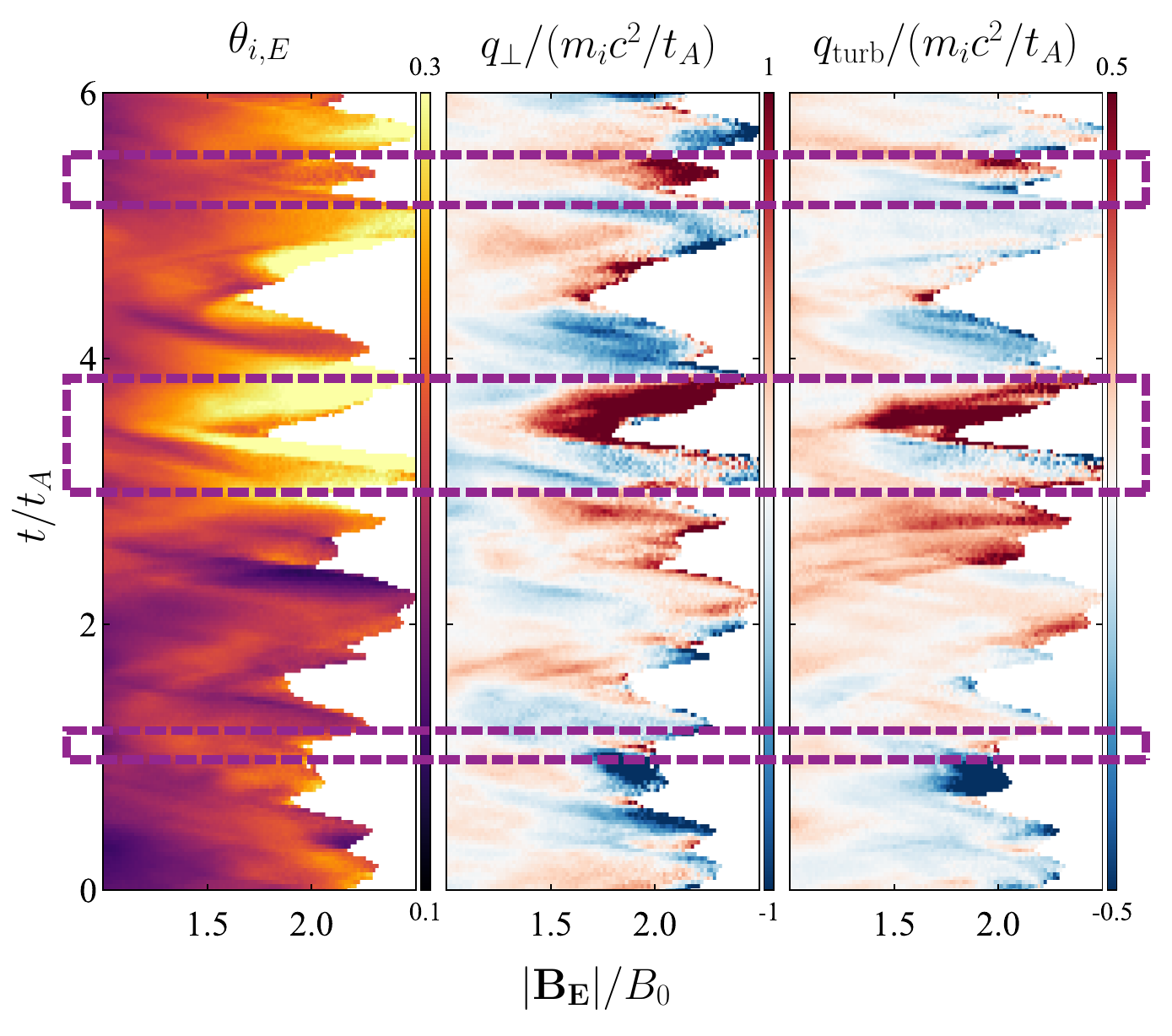}
	\caption{Long-term evolution of the ion temperature in the Eckart frame $\theta_{i,E}$, and the perpendicular gyrotropic and nongyrotropic work rates per ion, binned by $|\mathbf{B}|/B_0$}.
	\label{FigS5}
\end{figure}

Fig~\ref{FigS5} presents statistics of ion heating in strong magnetic-field structures over an extended interval of $6t_A$. Hot ion clumps appear intermittently in time: larger structures sustain heating for longer periods, whereas smaller structures can exhibit intense, localized bursts of energy dissipation, as illustrated by the boxed time intervals. Overall, ions undergo substantial heating within strongly magnetized structures through the combined action of the two heating mechanisms.

\end{document}